\documentclass{emulateapj}
\usepackage{natbib}
\begin{document}

\title{A Michelson-type Radio Interferometer for University Education}

\author{Jin Koda\altaffilmark{1},
James Barrett\altaffilmark{1}, 
Tetsuo Hasegawa\altaffilmark{2,3},
Masahiko Hayashi\altaffilmark{3},
Gene Shafto\altaffilmark{1},
Jeff Slechta\altaffilmark{1}, and
Stanimir Metchev\altaffilmark{1,4}}
\altaffiltext{1}{Department of Physics and Astronomy, Stony Brook University, Stony Brook, NY 11794-3800}
\altaffiltext{2}{National Astronomical Observatory of Japan, NAOJ Chile Observatory, Joaqu\'{\i}n Montero 3000 Oficina 702, Vitacura, Santiago 763-0409, Chile}
\altaffiltext{3}{National Astronomical Observatory of Japan, 2-21-1 Osawa, Mitaka, Tokyo 181-0015, Japan}
\altaffiltext{4}{Department of Physics and Astronomy, The University of Western Ontario, 1151 Richmond St, London, ON N6A 3K7, Canada}
\email{jin.koda@stonybrook.edu}

\submitted{Accepted for publication in American Journal of Physics, January 14, 2016} 

\begin{abstract}
We report development of a simple and affordable radio interferometer suitable as an educational laboratory experiment.
The design of this interferometer is based on the Michelson \& Pease stellar optical interferometer,
but operates at a radio wavelength ($\sim$11 GHz; $\sim 2.7$ cm); thus the requirement for
optical accuracy is much less stringent.
We utilize a commercial broadcast satellite dish and feedhorn. Two flat side mirrors slide on a ladder, providing baseline coverage.
This interferometer resolves and measures the diameter of the Sun, a nice daytime experiment
which can be carried out even in marginal weather (i.e., partial cloud cover).
Commercial broadcast satellites provide convenient point sources for comparison to the Sun's extended disk.
We describe the mathematical background of the adding interferometer, the design and development of the telescope
and receiver system, and measurements of the Sun. We present results from a students' laboratory report.
With the increasing importance of interferometry in astronomy, the lack of educational interferometers is an obstacle to training
the future generation of astronomers. This interferometer provides the hands-on experience needed
to fully understand the basic concepts of interferometry.
\end{abstract}

\maketitle 



\section{Introduction}\label{sec:intro}

The future of radio astronomy relies strongly on interferometers (e.g., ALMA, EVLA, VLTI, aperture masking techniques).
 From our experience at interferometer summer schools at the Nobeyama Radio Observatory and at the CARMA Observatory, we are convinced that hands-on experiments are critical to a full understanding of the concepts of interferometry. It is difficult, if not impossible, to obtain guaranteed access to professional interferometers for university courses. Therefore, we built a low-cost radio interferometer for the purpose of education and developed corresponding syllabi for undergraduate and graduate astronomy lab courses.

This experiment teaches the basic concept of interferometry using the technique developed by Michelson \& Peace in the early 20th century \citep{Michelson:1921uq}. They measured the diameter of Betelgeuse, one of the brightest stars in the sky, with a simple optical interferometer. Such optical interferometry needs high precision telescope optics. The same experiment becomes much easier  when measuring the diameter of the Sun at radio wavelength; the acceptable errors in the optics scale with the wavelength.

\begin{figure*}
\epsscale{1.0}
\plotone{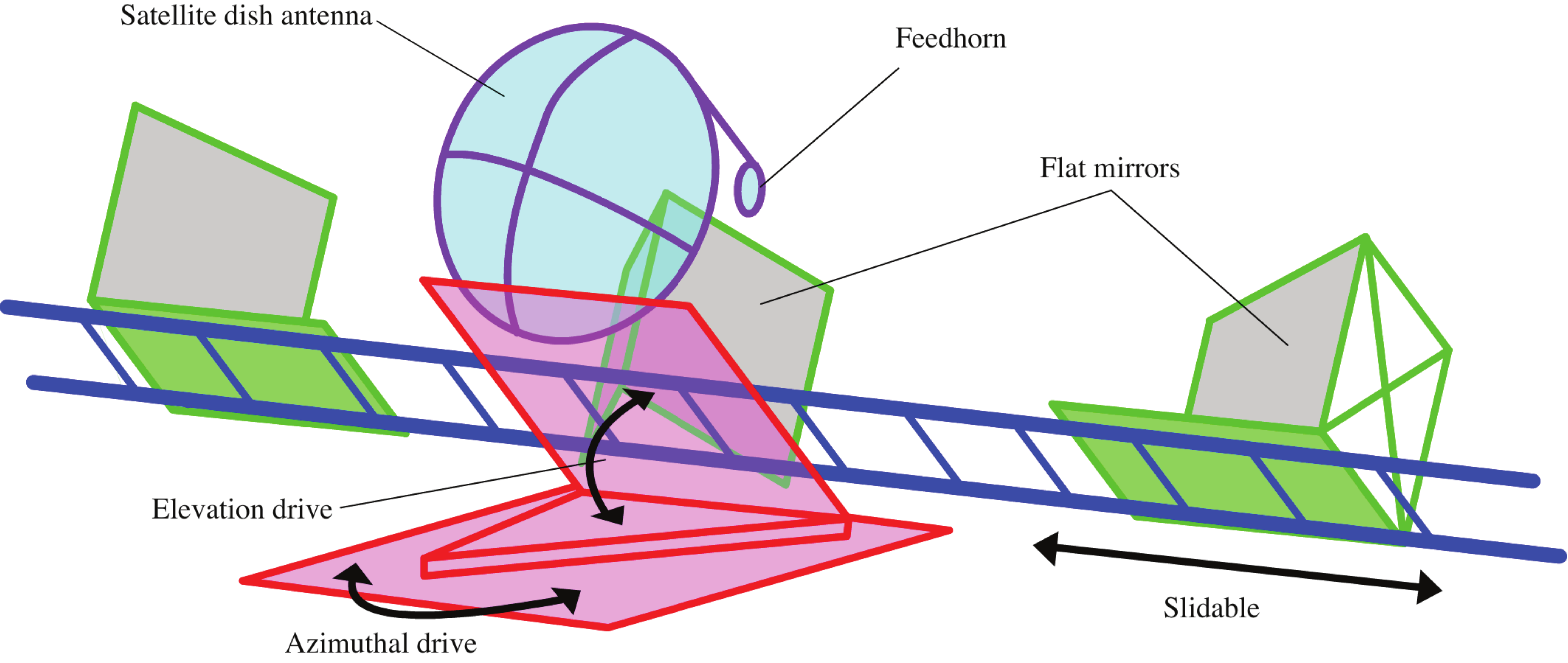}
\caption{Conceptual sketch of the Michelson radio interferometer.}
\label{fig:sketch}
\end{figure*}

Figure \ref{fig:sketch} shows a conceptual sketch of the Michelson radio interferometer for education. This type of interferometer, adding signals instead of multiplying them, is called an adding interferometer. We discuss the mathematical background of the adding interferometer in \S \ref{sec:math}, design and development of the telescope and receiver system in \S \ref{sec:instrument}, telescope setup and measurements in \S \ref{sec:setup}, and results from a students' lab report in \S \ref{sec:results}.
What we present here is only one realization of the concept. Creative readers could modify any part to meet the educational needs and constraints at their own institutions. For example, the astronomical measurement, the construction and tests of of telescope, receiver system, and other components can be separate lab projects.

The best known Michelson interferometer is the one used for the Michelson-Morley experiment \citep{Michelson:1887lr}.
It is one of the most important classical experiments taught in both lecture and laboratory courses \citep{Wolfson99, Melissinos03, Serway13, Bennett13}.
Many studies and applications have appeared in this journal\citep{Fang13,Rudmin80,Matthys82,daCosta88,Diamond90,Mellen90,Norman92,Belansky93,Kiess96,Nachman97,Fox99},
and recently the Michelson interferometer is being applied to the detection of gravitational waves \citep{Kuroda:1999rt, Abbott:2009yq, Harry:2010vn}.
The Michelson stellar interferometer is an application of the same physical concept of interference,
in this case, to a light source in the sky.

The theoretical basis of the Michelson stellar interferometer was already established in the Michelson and Peace's original
work \citep{Michelson:1921uq} and has been used in radio interferometry, especially in its early history \citep{Pawsey55, Steinberg63, Christiansen85, Wilson13}.
This adding interferometer is the type used in modern astronomy at optical and near-infrared wavelengths \citep{Shao:1992fk, Quirrenbach:2001lr}
though modern radio interferometers are of a different type, multiplying signals instead of adding them \citep{Taylor:1999ab, Thompson:2007fk}.
For educational purposes, some studies in this journal showed that the concept of the stellar interferometer
could be demonstrated in an indoor laboratory setup using an artificial light source  \citep{Pryor59,Illarramendi14}.
In professional optical astronomy, the technique is now being applied for advanced research \citep{Shao:1992fk, Quirrenbach:2001lr, Monnier:2003uq}.


\section{Mathematical Background}\label{sec:math}

The mathematical basis of the stellar interferometer was presented in Michelson and Peace's
original work\citep{Michelson:1921uq} and can be found in textbooks \citep{Pawsey55, Steinberg63, Christiansen85, Wilson13}.
Here we describe the basic equations at a mathematical level that college students can follow.

We start from the geometric delay calculation (\S \ref{sec:delay}) and explain the total power,
the parameter that we measure, in \S \ref{sec:totalpower}.
We will show an example of how a point source (i.e., a commercial broadcast satellite)
appears in \S \ref{sec:pointsource}.
We will then discuss the case of an extended source.
We prove that an interferometer measures Fourier components
and define visibility in \S \ref{sec:visibility}.
We will explain how visibility is measured with our interferometer,
and how the Sun's diameter is derived in \S \ref{sec:measure}.

\subsection{Geometric Delay}\label{sec:delay}

\begin{figure}
\epsscale{1.1}
\plotone{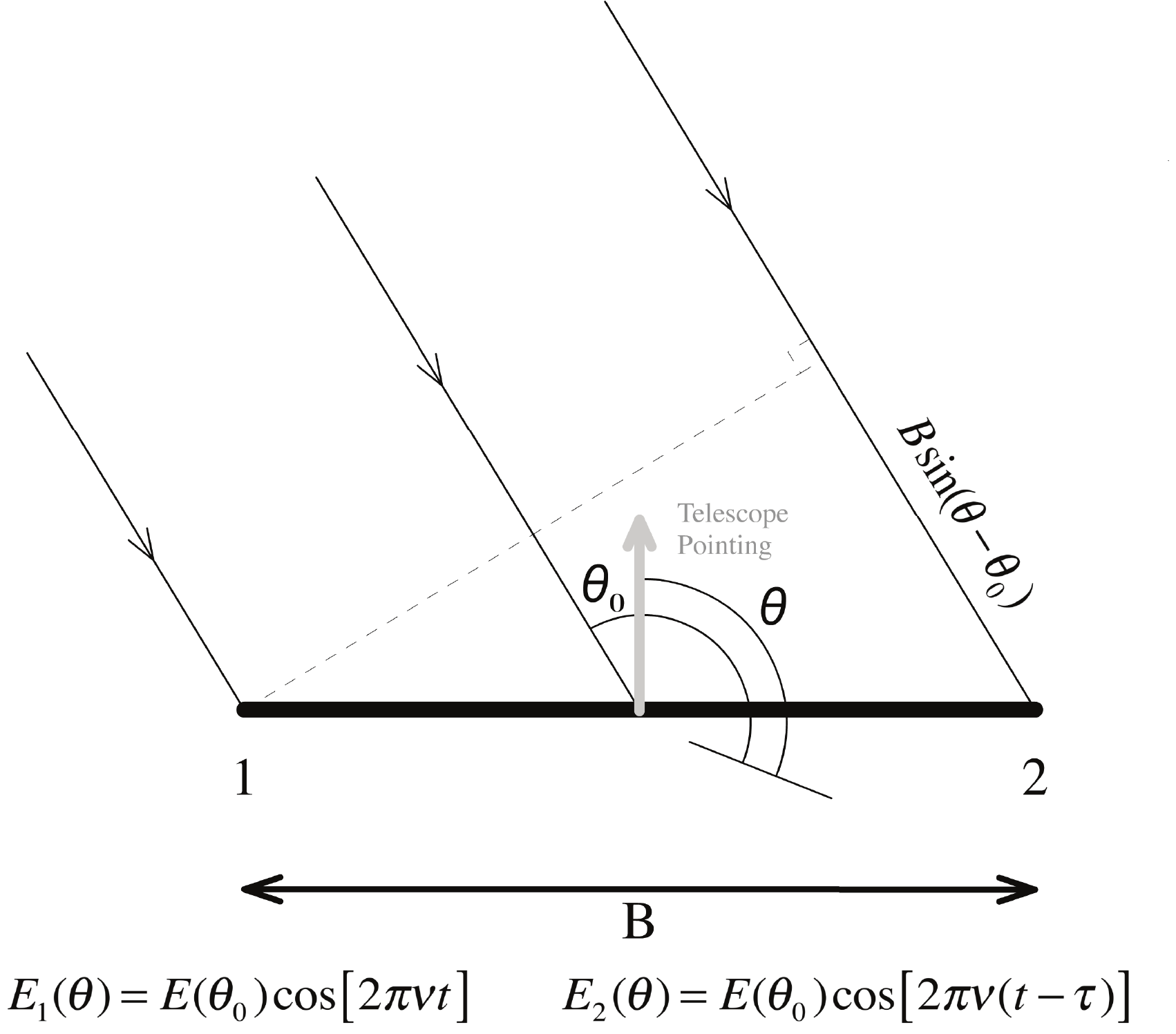}
\caption{Schematic illustration of signal detection with two detectors separated by the baseline length $B$.
The direction of the telescope pointing is $\theta$ and that to an object in sky is $\theta_0$, both from an arbitrary
origin.}
\label{fig:schemsignal}
\end{figure}

Interferometers mix signals received at two different positions (position 1 \& 2 in Figure \ref{fig:schemsignal}).
In our radio interferometer, the signals that arrive at the two side mirrors (Figure \ref{fig:sketch}) are guided to
the antenna and mixed.
The separation between the two mirrors, called baseline length $B$, causes a time delay
$\tau$ in the arrival of the signal at position 2 because of the geometry (Figure \ref{fig:schemsignal}).
Using the angles of the telescope pointing $\theta$ and to an object in the sky $\theta_0$, a simple
geometric calculation provides the delay,
\begin{equation}
\tau = \frac{B \sin (\theta - \theta_0)}{c} \sim \frac{B (\theta-\theta_0)}{c} \label{eq:delay}
\end{equation}
where $c$ is the speed of light.
We used the small angle approximation, $\sin(\theta-\theta_0)\sim \theta-\theta_0$, since most astronomical objects
have a small angular size.


\subsection{Total Power}\label{sec:totalpower}

Radio signals are electromagnetic radiation and can be described in terms of an electric field $E$ and a magnetic field $B$.
For simplicity, we consider only the electric field $E$ in the following calculations (but this simplification does not
lose the generality of the discussion).  If we define the radio signal at frequency $\nu$ that is detected
at position 1 (or reflected if a mirror is there) at time $t$ as,
\begin{equation}
E_1(t) = E(\theta_0) \cos[2\pi \nu t],
\end{equation}
the signal that is detected at position 2 at the same time is,
\begin{equation}
E_2(t) = E(\theta_0) \cos[2\pi \nu (t-\tau)],
\end{equation}
because of the geometric delay $\tau$.

An adding interferometer adds the two signals and measures total power of the two.
The total electric field is
\begin{equation}
E_{\rm tot}(t) = E_1(t) + E_2(t). \label{eq:etot1}
\end{equation}
The radio frequency $\nu$ is typically large compared to a data sampling rate.
Hence, the total power $P(\theta)$, detected by a receiver, is a time average (or integration).
Using the notation $< ...>$ for the time average, we obtain
\begin{eqnarray}
P(\theta)&=&\left<E_{\rm tot}^2(\theta)\right>\\
             &=&\left<E^2(\theta_0)\left(  \cos[2\pi \nu t] +  \cos[2\pi \nu (t-\tau)] \right)^2\right> \\
             &=&\left<E^2(\theta_0)\left(  \cos^2[2\pi \nu t] + \cos^2[2\pi \nu (t-\tau)]  \right. \right. \nonumber \\
             && \left. \left. + 2  \cos[2\pi \nu t]\cos[2\pi \nu (t-\tau)]\right)\right>\label{eq:im2}\\
            & =&E^2(\theta_0)[1+\cos(2\pi\nu\tau)] \label{eq:im3} 
\end{eqnarray}
In going from eq (\ref{eq:im2}) to (\ref{eq:im3}) we used the transformations: $\cos^2A = (\cos 2A+1)/2$ for the first and third terms
and $2 \cos A \cos B = \cos(A+B) + \cos (A-B)$ for the second term.
In addition, because of the high frequency, $\nu$, all terms with $\left<\cos(*\nu t)\right>$, $\left <\sin(*\nu t) \right>$, etc,
vanish when time averaged, and only the terms with no $t$ dependence remain.
Using equation (\ref{eq:delay}) with the small angle approximation, this becomes
\begin{equation}            
P(\theta)=E^2(\theta_0)[1+\cos(2\pi B_\lambda(\theta-\theta_0))] \label{eq:power2}
\end{equation}
where $B_\lambda \equiv B/\lambda$ is a normalized baseline length and $\lambda$ is the wavelength ($\lambda =c/\nu$).

Equation (\ref{eq:power2}) can be generalized for an extended object as
\begin{equation}
P(\theta)=\int{\cal E}(\theta_0)d\theta_0[1+\cos(2\pi B_\lambda(\theta-\theta_0))], \label{eq:power3}
\end{equation}
where ${\cal E}(\theta_0)$ is an intensity/energy density distribution of the object.
Our adding interferometer measures  $P(\theta)$; we slew the telescope across the object
in the azimuthal direction and obtain  {\it fringes}, i.e., variations in the power as a function of $\theta$.

\subsection{Point Source}\label{sec:pointsource}

The energy density of a point source is a $\delta$-function at the position of the object
$\theta_0 = \theta_{\rm c}$. By adopting the coordinate origin to make $\theta_{\rm c} = 0$,
it is
\begin{equation}
{\cal E}(\theta_0)={\cal E}_0\delta(\theta_0).
\end{equation}
Combining with eq. (\ref{eq:power3}), we obtain
\begin{equation}
P(\theta)={\cal E}_0[1+\cos(2\pi B_\lambda(\theta-\theta_0))].\label{eq:power_point}
\end{equation}
As we sweep the telescope from one side of the object to the other,
we should see a sinusoidal power response as a function of $\theta$.

\begin{figure}
\epsscale{1.2}
\plotone{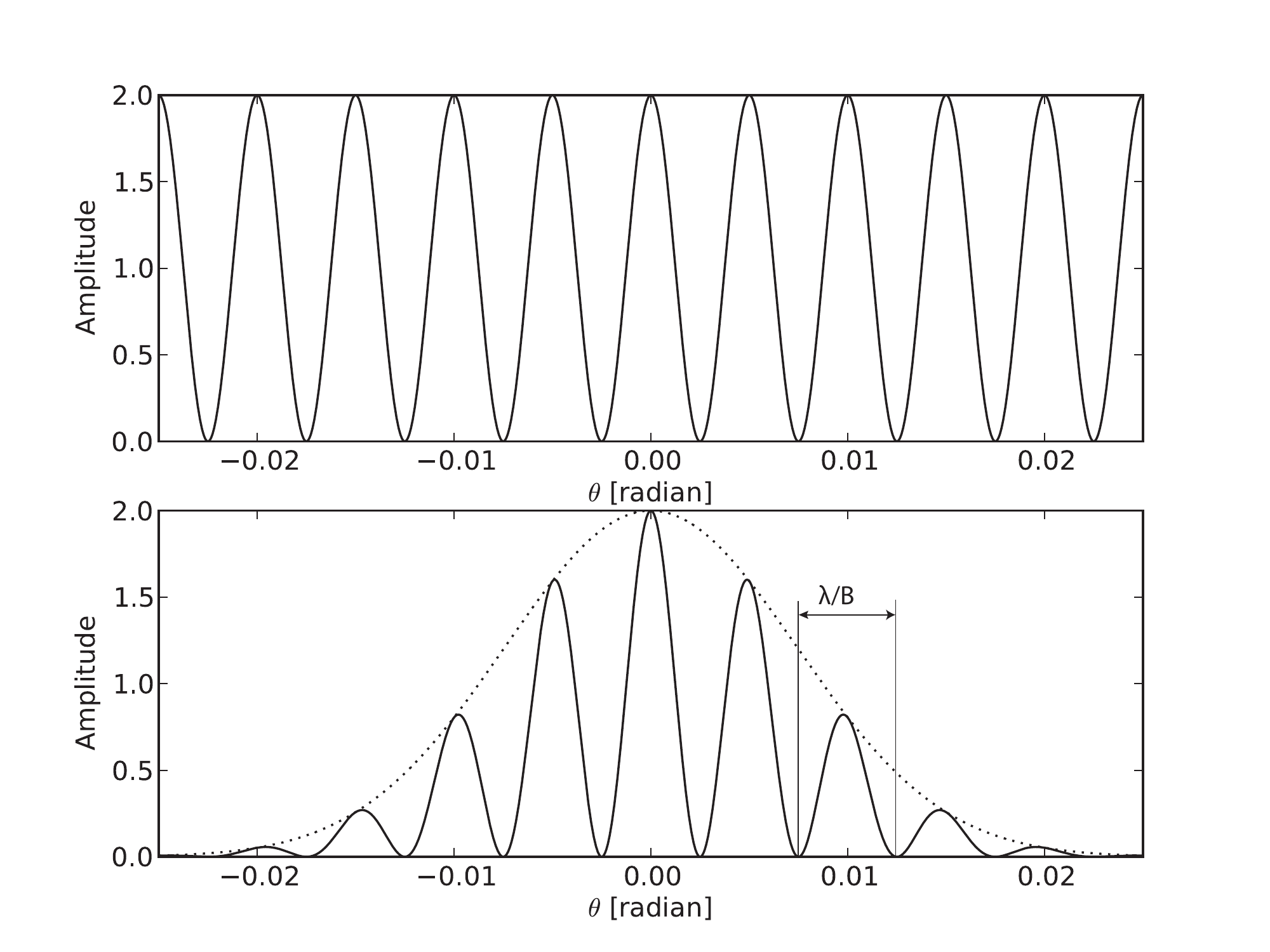}
\caption{Example plots of the total power as a function of telescope pointing $\theta$ in the case of a point source.
{\it Top:} Fringe pattern (eq. \ref{eq:power_point}).
{\it Bottom:} Fringe pattern attenuated by the telescope beam pattern. The dotted-line is a Gaussian beam pattern with a FWHM of 1 degree.}
\label{fig:satellite}
\end{figure}

Figure \ref{fig:satellite} ({\it top}) shows the theoretical fringe pattern from a point source.
Our satellite dish (and any other radio telescope) has a directivity; its response
pattern tapers off away from the center.
The pattern that we actually obtain is attenuated by the dish response pattern
(beam pattern) as shown in Figure \ref{fig:satellite} ({\it bottom}).
Commercial broadcast satellites are very small in angle and approximate point sources.

Fringe measurements are useful in determining the baseline length $B_\lambda$.
The total power is zero when the normalized baseline is $B_\lambda(\theta-\theta_0) = n+1/2$,
where $n$ is an integer.
The separation between adjacent null positions is $\delta \theta = 1/B_\lambda = \lambda/B$.


\subsection{Extended Source and Visibility}\label{sec:visibility}

An astronomical object is often extended.
In general, an interferometer measures the Fourier transform of the energy
density distribution ${\cal E}(\theta_0)$. Here we prove this.

From eq. (\ref{eq:power3}) we define the visibility $V_0(B_\lambda)$ as follows:
\begin{eqnarray}
P(\theta)
&=& \int {\cal E}(\theta_0) d\theta_0 \nonumber\\
& & +\int {\cal E}(\theta_0)\cos(2\pi B_\lambda(\theta-\theta_0)) d\theta_0\\
&\equiv&S_0[1+V(\theta, B_\lambda)], \label{eq:power4}
\end{eqnarray}
where
\begin{equation}
S_0\equiv \int {\cal E}(\theta_0) d\theta_0
\end{equation}
and 
\begin{eqnarray}
V(\theta, B_\lambda)&\equiv&{{1}\over{S_0}}\int {\cal E}(\theta_0) \cos[2\pi B_\lambda(\theta-\theta_0)] d\theta_0\\
&=&{{1}\over{S_0}}\left[\cos(2\pi B_\lambda\theta)\int {\cal E}(\theta_0)\cos(2\pi B_\lambda\theta_0) d\theta_0 \right. \nonumber\\
& &\left. + \sin(2\pi B_\lambda\theta)\int {\cal E}(\theta_0)\sin(2\pi B_\lambda\theta_0) d\theta_0 \right]  \\
&\equiv& V_0(B_\lambda) \cos[2\pi B_\lambda(\theta-\Delta\theta)]. \label{eq:vis1}
\end{eqnarray}
Here, the visibility $V_0(B_\lambda)$ and the phase shift $\Delta \theta$ are defined as
\begin{eqnarray}
V_0(B_\lambda)\cos(2\pi B_\lambda\Delta\theta)&=&{{1}\over{S_0}}  \int {\cal E}(\theta_0)\cos(2\pi B_\lambda\theta_0)d\theta_0, \\
V_0(B_\lambda)\sin(2\pi B_\lambda\Delta\theta)&=&{{1}\over{S_0}}  \int {\cal E}(\theta_0)\sin(2\pi B_\lambda\theta_0)d\theta_0,
\end{eqnarray}
which lead to
\begin{equation}
V_0(B_\lambda) =e^{i2\pi B_\lambda\Delta\theta} 
{{1}\over{S_0}} 
\int {\cal E}(\theta_0) e^{- i2\pi B_\lambda\theta_0}d\theta_0.
\end{equation}
The first term $e^{i2\pi B_\lambda\Delta\theta}$ is a phase shift $\Delta \theta$ of a complex visibility.
The visibility amplitude is therefore
\begin{equation}
\left|V_0(B_\lambda)\right| = \left|{{1}\over{S_0}} \int {\cal E}(\theta_0) e^{- i2\pi B_\lambda\theta_0}d\theta_0 \right|.
\end{equation}
This is a Fourier component of the object ${\cal E}(\theta_0)$ at a baseline length of $B_\lambda$.
The inverse $1/B_\lambda$ is the angular size of the Fourier component in radians.
Observations at long baseline lengths detect structures of small angular size (i.e., Fourier components corresponding to small angular structures),
while those at short baselines capture structures of large angular size.

\begin{figure}
\epsscale{1.2}
\plotone{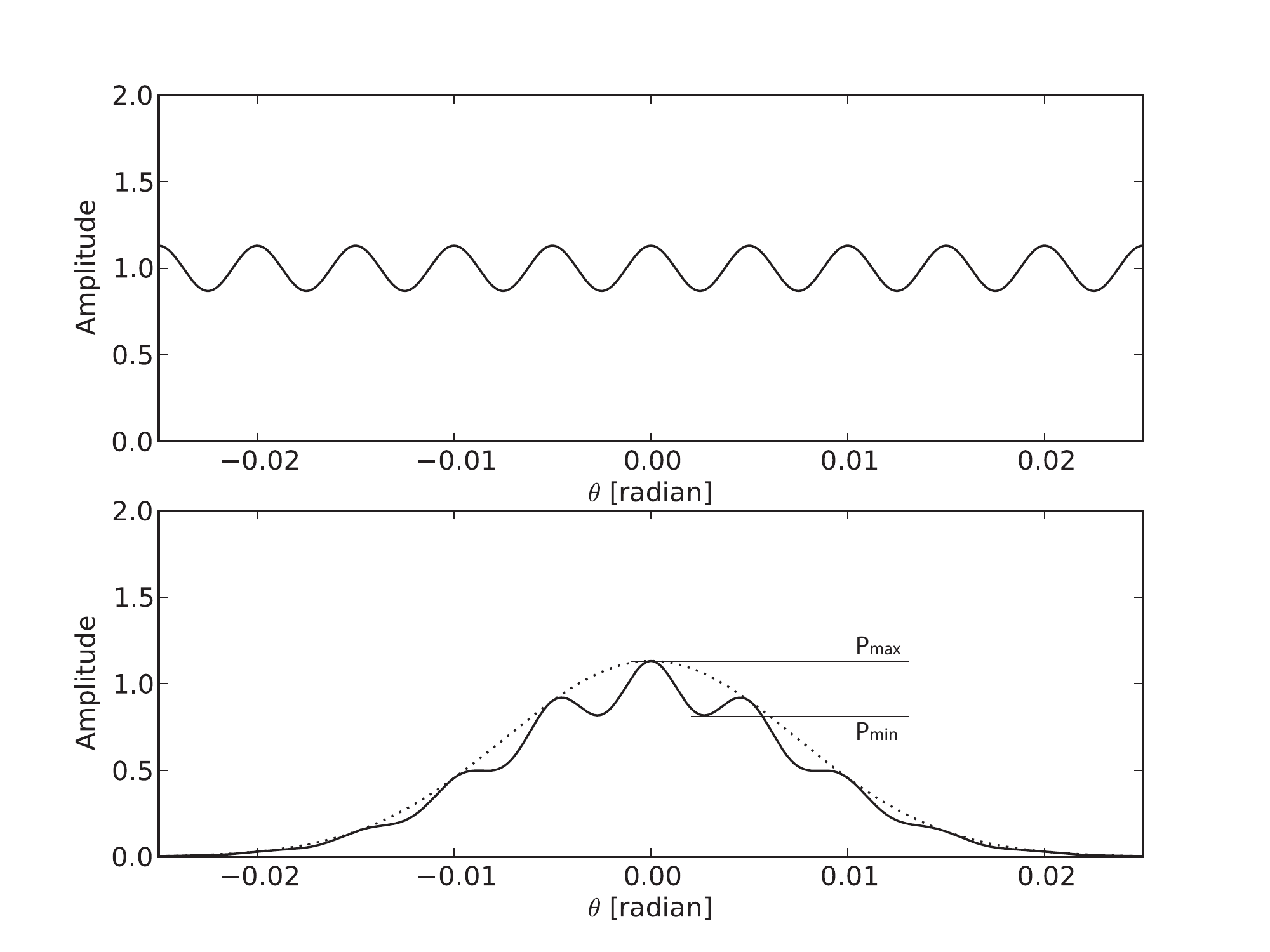}
\caption{Example plots of the total power as a function of telescope pointing $\theta$ in case of disk (like the Sun).
{\it Top:} Fringe pattern (eq. \ref{eq:power_point}).
{\it Bottom:} Fringe pattern attenuated by the telescope beam pattern. The dotted-line is a Gaussian beam pattern with a FWHM of 1 degree.}
\label{fig:sun}
\end{figure}

Figure \ref{fig:sun} ({\it top}) shows the theoretical fringe pattern for the top-hat function (e.g., the Sun's disk in 2-dimensions).
The pattern is also attenuated by the beam pattern (Figure  \ref{fig:sun} {\it bottom}).


\subsection{Visibility Measurements and Sun's Diameter}\label{sec:measure}

We measure $P(\theta)$ and calculate the visibility amplitude $|V_0(B_\lambda)|$.
From eqs. (\ref{eq:power4}) and (\ref{eq:vis1}), we have
\begin{equation}
P(\theta)=S_0\left[1+V_0(B_\lambda)\cos\left[2\pi B_\lambda(\theta-\Delta\theta)\right]\right]
\end{equation}

Figure  \ref{fig:sun} ({\it bottom}) is what we see toward the Sun -- we sweep across the Sun
by slewing the telescope in the azimuthal direction (i.e., changing $\theta$).
The fringe pattern is attenuated by the antenna response pattern,
but we assume that the antenna response
is approximately constant around the peak of the response pattern.
The maximum and minimum powers of the sinusoidal curve (see Figure  \ref{fig:sun} {\it bottom}) are
\begin{eqnarray}
P_{\rm max}&=&S_0[1+V_0(B_\lambda)] \\
P_{\rm min}&=&S_0[1-V_0(B_\lambda)].
\end{eqnarray}
From these, we calculate
\begin{equation}
\left| V_0(B_\lambda) \right| ={{P_{\rm max}-P_{\rm min}}\over{P_{\rm max}+P_{\rm min}}}.
\end{equation}
This is the visibility amplitude at a baseline length of $B_\lambda$.

The two side mirrors slide on the ladder in Figure \ref{fig:sketch} and change the baseline length.
We repeat measurements of $\left| V_0(B_\lambda) \right|$ at different baseline lengths
and make a plot of $\left| V_0(B_\lambda) \right|$ as a function of $B_\lambda$.
$\left| V_0(B_\lambda) \right|$ is a Fourier component of ${\cal E}(\theta_0)$; therefore,
we should see the Fourier transformation of the emission distribution in the plot.

\begin{figure}
\epsscale{1.2}
\plotone{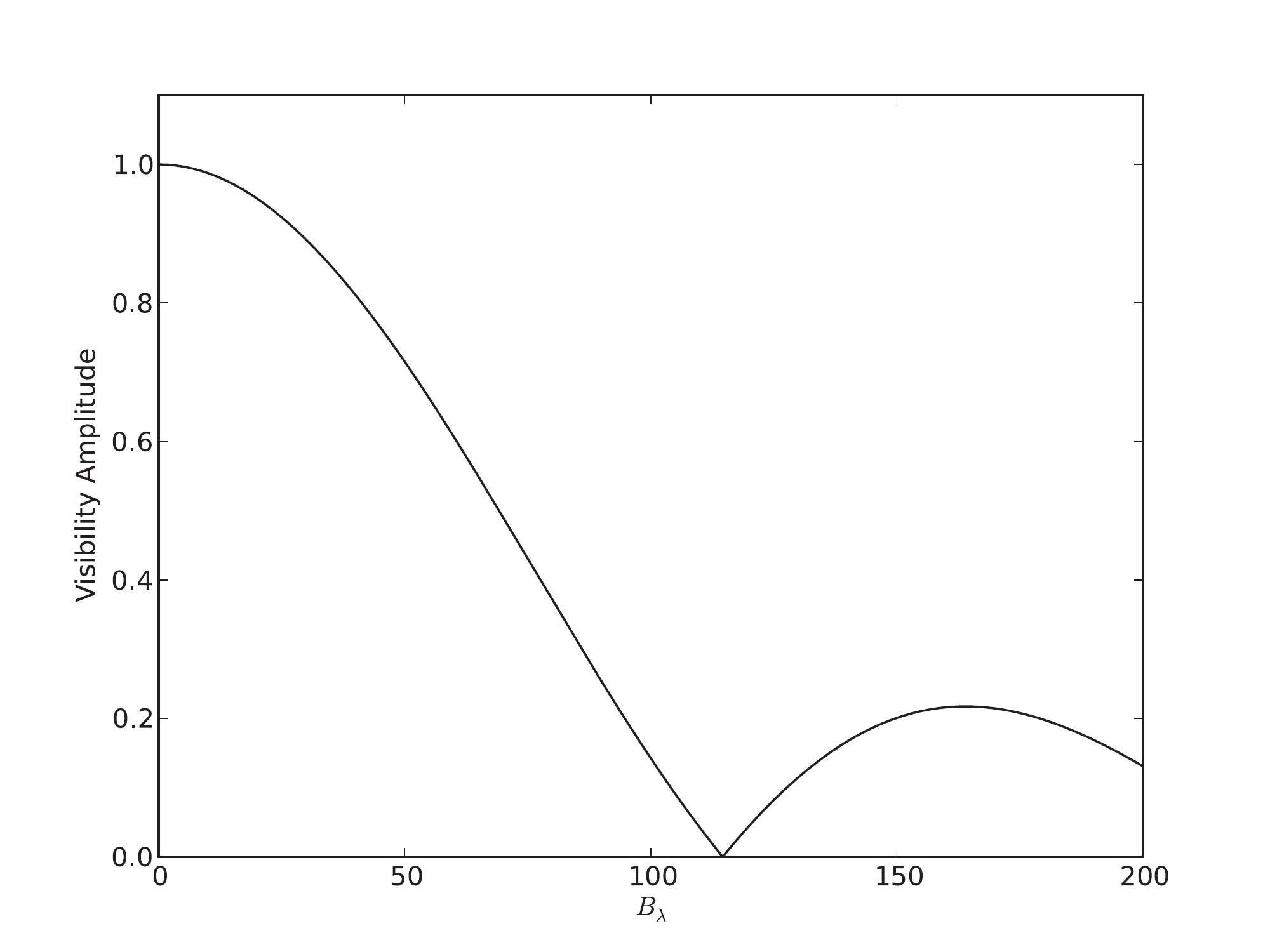}
\caption{Visibility amplitude as a function of baseline length in the case of a disk.}
\label{fig:visibility}
\end{figure}

The Sun's  ${\cal E}(\theta_0)$ can be approximated as a top-hat function.
Assuming the Sun's diameter is $\alpha$, it is
\begin{equation}
{\cal E}(\theta_0) =
\left\{
\begin{array}{rl}
1, & \text{if } |\theta_0| < \alpha/2 \\
0, & \text{otherwise}
\end{array}
\right.
\end{equation}
The Fourier transform is
\begin{equation}
\left| V_0(B_\lambda) \right| = \frac{\sin(\pi B_{\lambda} \alpha)}{\pi B_{\lambda}}.
\end{equation}
This is a {\it sinc} function (Figure \ref{fig:visibility}).
By fitting, we determine the parameters of this {\it sinc} function,
which can be translated to the diameter of the Sun $\alpha$.
(This is a 1-dimensional approximation of the Sun's shape.
A more ambitious exercise would be to use a more accurate treatment of its 2-dimensional shape.)


\section{Instruments}\label{sec:instrument}

We describe the construction of the telescope and receiver system.
The budget is often the main limitation in the development of student lab experiments.
Hence, we utilized low-cost parts and materials and
used a commercial broadcast satellite dish and feedhorn operating at radio X-band.
The system was constructed in our machine and electronics shops.
Fabrication of the components could be offered as a student lab projects.

\subsection{Telescope and Optics}
\begin{figure*}
\epsscale{1.0}
\plotone{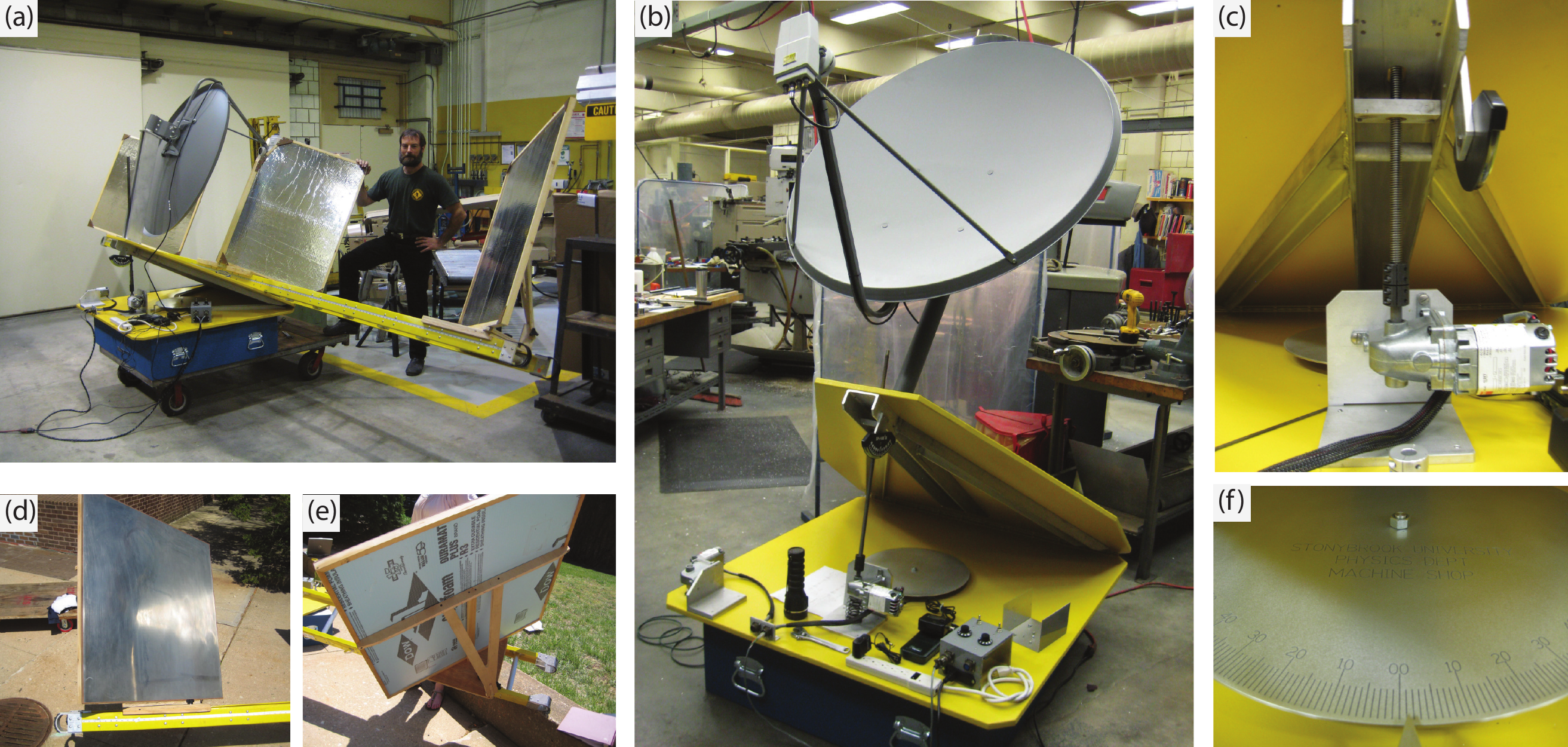}
\caption{
Photographs of the telescope.
(a) Overall view.
(b) Mount structure. The blue box at the bottom (with handles) and yellow plates are made of wood. The entire yellow part rotates in the azimuthal direction on the blue box. The two yellow plates are attached with hinges, and the top plate moves up to change the elevation angle.
The telescope is shown in a "single-dish" mode, and the dish would be rotated by $180\deg$ for an "interferometer" experiment.
(c) Support structure. The aluminum frame supports the telescope. A screw rod and elevation drive motor are also visibile.
(d) Side mirror from the front side. Kitchen aluminum foil is thicker than the required skin depth, but we glued a thin aluminum plate instead, as student-proofing.
(e) Side mirror from the backside. It's supported by a wood frame.
(f) Protoractor to measure the azimuthal angle of the telescope. }
\label{fig:telescope}
\end{figure*}

Figure \ref{fig:sketch} shows the design of the Michelson stellar radio interferometer.
Radio signals from the Sun hit two flat mirrors at the sides and are reflected to a satellite dish
antenna by the central flat mirrors.
The signals from the two sides are mixed as detected.
Figure \ref{fig:telescope} shows photos of the telescope.
It was built with mostly commercial products and materials.
A broadcast satellite dish and feedhorn (blue in Figure \ref{fig:sketch};
Figure \ref{fig:telescope}a,b) operates at a frequency of $\nu\sim11$ GHz ($\lambda \sim 2.7$ cm
in wavelength).
The required accuracy of optics at this wavelength is about $\sim$3-5 mm,
which is relatively easy to achieve with flat mirrors (without curvature).

The flat mirrors (green in Figure \ref{fig:sketch}) are made of fiberboard
with wooden framing structures (Figure \ref{fig:telescope}e).
The mirror surfaces are all angled $45\deg$ from the optical path.
We originally covered their surfaces with kitchen aluminum foil,
which has an appropriate thickness with respect to the skin depth ($\sim 0.8\mu m$)
at the operating wavelength (reflectivity $\sim 96$\% from our lab measurements).
Later, we replaced it with thin aluminum plates as student-proofing (Figure \ref{fig:telescope}d).
The two side mirrors slide on a ladder to change the baseline length.

The azimuth-elevation mount structure is made with plywood (red in Figure \ref{fig:sketch}
and blue and yellow in Figure \ref{fig:telescope}).
The azimuthal and elevation axes are driven with motors (Figure \ref{fig:telescope}c),
which are controlled by a paddle (i.e., handset in Figure \ref{fig:telescope}b).
The protractor (Figure \ref{fig:telescope}f) is placed at the center of the bottom mount plate
(yellow in Figure \ref{fig:telescope}b) for measurement of the azimuthal angle of the telescope.
Figure \ref{fig:telescope}a shows the whole structure of the telescope.
A metal pole is mounted perpendicular to the top mount plate (Figure \ref{fig:telescope}b)
and aluminum frame (Figure \ref{fig:telescope}c), and supports the dish.
Note that the pole should be perpendicular, which makes the pointing adjustment
easier as discussed later.

The azimuthal rotation is facilitated by greased handcrafted ball bearings in circular
grooves around the azimuth shaft on the base (blue in Figure \ref{fig:telescope}a,b - below the yellow structure)
and on the bottom mount plate (yellow).

Sweeping across the Sun in azimuth permits fringe measurements.
This telescope can be converted to a single-dish telescope
by flipping the satellite dish by 180 degrees around the metal pole (see Figure \ref{fig:telescope}b).
Single-dish and interferometer measurements can be easily made and compared,
which is essential for appreciation of the high angular resolution possible with the interferometer.

Table \ref{tab:telescope} lists the commercial product parts that we purchased.
The other parts, mostly the support structure, are made in the machine shop.


\begin{table*}[h]
\centering
\caption{Commercial Products Purchased for Telescope Mount}
\begin{ruledtabular}
\begin{tabular}{ccccccc}
No & Description  & Quantity & Manufacturer & Part No. & Vendor & Price \\
\hline
1 & Manhole Ladder 16 ft                & 1 & Werner      &  M7116-1    & Lowe's & \$226 \\
2 & Motor                                          & 2 &  Dayton      &  1LPZ7    &  Walmart   & \$248 \\
3 & Lev-O-Gage                                             & 1 & Sun Company, Inc.         &  NWH-0152-1003  & opentip.com & \$18 \\
\end{tabular}
\end{ruledtabular}
\label{tab:telescope}
\end{table*}

\begin{table*}[h!]
\centering
\caption{Purchased Receiver Parts}
\begin{ruledtabular}
\begin{tabular}{ccccccc}
No & Description  & Quantity & Manufacturer & Part No. & Vendor & Price \\
\hline
1 & 1-Meter Satellite Dish                                    & 1 & WINEGARD      & DS-3100     & Solid Signal & \$90 \\
2 & Quad Polar LNBF                                          & 1 & INVACOM       & QPH-031     & SatPro.tv    & \$55 \\
3 & Power Inserter                                           & 1 & PDI           & PDI-PI-1    & Solid Signal & \$2 \\
4 & 75-50 Ohm Adaptor & 1 & PASTERNACK    & PE7075      & Pasternack   & \$83 \\
5 & Amplifier 50$1/2$ 0.5 to 2.5 GHz                             & 1 & Mini-Circuits & ZX60-2534M+ & Mini-Circuits & \$65 \\
6 & Attenuator SMA 3GHz 50 Ohm 10db                          & 1 & Crystek       & CATTEN-0100 & Digi-Key     & \$19 \\
7 & Attenuator SMA 3GHz 50 Ohm 6db                           & 1 & Crystek       & CATTEN-06R0 & Digi-Key     & \$19 \\
8 & Bandpass Filter 1350 to 1450 MHz                         & 1 & Mini-Circuits & ZX60-2534M+ & Mini-Circuits & \$40 \\
9 & Square-Law Detector 1.0-15.0 GHz                                  &  1 &  Omni Spectra &  Model 20760 & eBay  & \$30 \\
10 & 5X OP-Amp                                                & 1 & Custom Built\footnote{This component could be simply some batteries that provide the voltage of $\sim5~$V.}  &             &              & \$20 \\
11 & IC Buck Converter Mod 5.0V SIP3                          & 1 & ROHM          & BP5277-50   & Digi-Key     & \$8 \\
12 & Box Aluminum 4'$\times$6'$\times$10' (HWD)                            & 1 & LMB Heeger    & UNC 4-6-10  & DigiKey     & \$45 \\
13 & 0-5V Analog Meter 4\"                                    & 1 & Salvaged      &             &              & \$0 \\
14 & Data Converter \& Collection                               & 1 & Vernier        & LabPro      & Vernier      & \$220
\end{tabular}
\end{ruledtabular}
\label{tab:receiver}
\end{table*}


\subsection{Receiver System}

\begin{figure*}
\epsscale{1.1}
\plotone{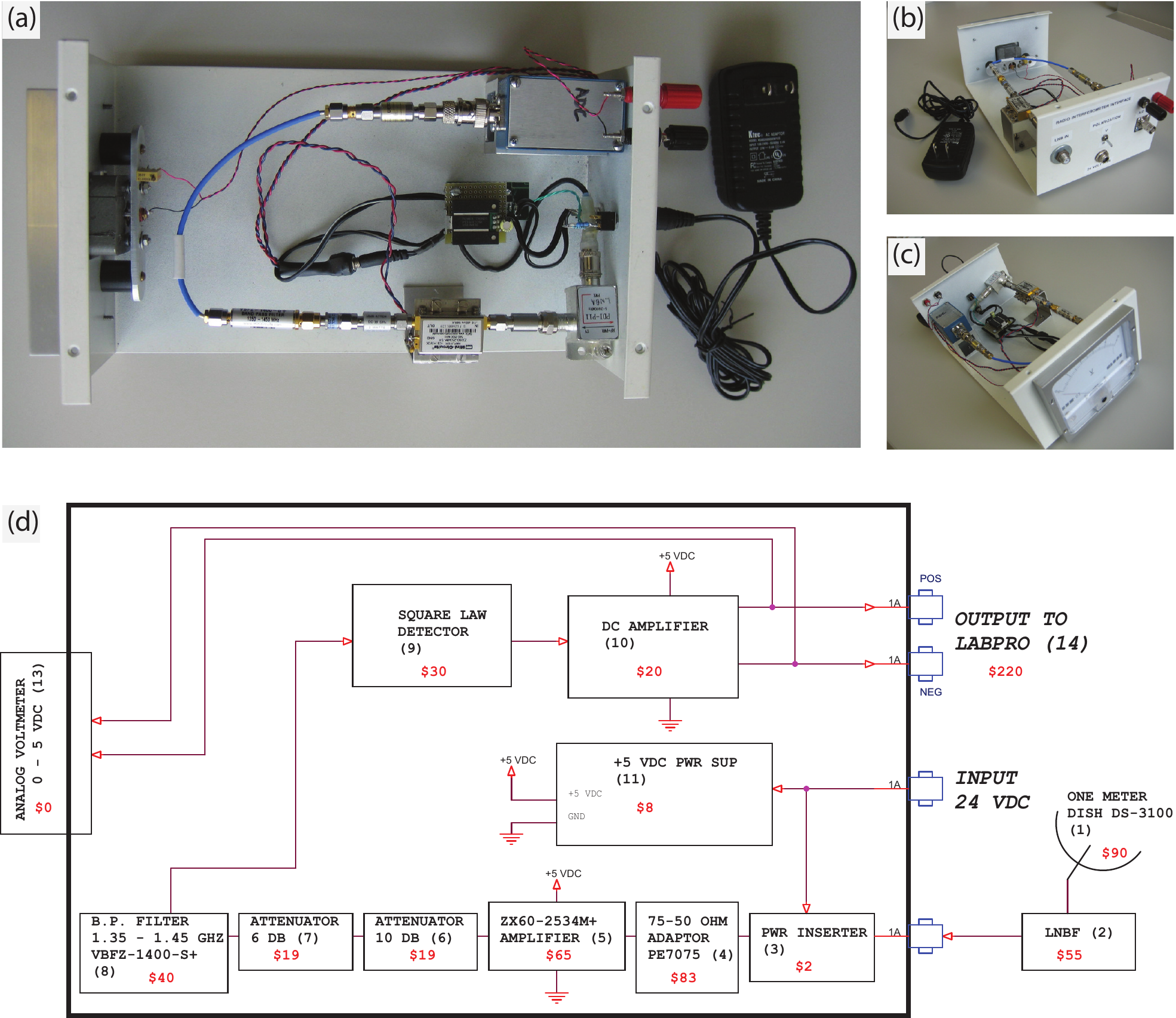}
\caption{Photographs and schematic of the receiver.
(a) Interior of receiver box. Most components are commercial.
(b) Front side of the receiver box. Two critical plugs are for an input from the feedhorn and
output to the LabPro (commercial analog/digital converter often used in physics lab courses, which outputs
to a computer through a USB connection).
(c) Back side. We installed an analog voltage meter, so that signal detection can be easily checked during observations.
(d) Schematic diagram of receiver components.}
\label{fig:receiver}
\end{figure*}

The signal detection system in radio astronomy is a series of electronic components.
Figure \ref{fig:receiver} shows the design and photos of the receiver.
Again, these are mostly commercial products.

Signals from the sky are at too high a frequency ($\sim 11$ GHz) to be handled electronically.
Hence the Low Noise Block Feedhorn (LNBF) down-converts the frequency
to a lower frequency, called the intermediate frequency (IF; 950-1950MHz), by mixing
the sky signal with a reference signal at a slightly-offset frequency and producing a signal at the beat frequency
of the sky and reference signals.
This is called heterodyne receiving. The LNBF works as a heterodyne mixer.

Figure \ref{fig:receiver} shows the flow of signal.
In sequence, an amplifier, two attenuators, and bandpass filter adjust the signal amplitude
to the input range of a square-law detector.
We combined two commercially available attenuators to achieve the desired attenuation of $\sim16$ db.
A filter with a 100 MHz width narrows the frequency range,
since the bandwidth of the IF (1GHz at the operating frequency of $\sim$11 GHz) is too broad
for detection of null fringes in interferometry.
Output from the detector is then amplified to the whole dynamic range of the analog-to-digital (A/D)
converter. We assembled all these components inside a metal box for protection.
A power supply is also in the box, providing the power to the LNBF and amplifiers.

The output from the receiver box goes to a commercial LabPro A/D convertor.
The LabPro is connected via USB to, and controlled by, a laptop computer with LabPro
software installed. It takes care of time integration and sampling rate for voltage measurements.

Table \ref{tab:receiver} lists the electronics components that we purchased.
The square-law detector (Schottky diode detector) was purchased through eBay,
and similar devices seem almost always on sale there.
We then found and purchased the amplifier and attenuators to adjust the signal voltage amplitude
to adjust the input range of the detector and the output range of the LNBF when the telescope is
pointing toward the Sun and satellites.


\section{Setup and Measurements}\label{sec:setup}

\subsection{Setup}
The mount structure, ladder, and mirrors of the telescope (Figure \ref{fig:telescope})
are detached when it is stored in our physics building.
We move them with a cart to the front of the building and assemble them there
on the morning of experiment.
We make sure that the flat mirrors are angled at $45\deg$ with respect to the optical path
and $90\deg$ vertically, using a triangle.
We then attach the ladder and mirrors to the mount structure using clamps mounted on the structure.

The electronic components are also connected: the signal from
the feedhorn goes to the receiver (Figure \ref{fig:receiver}), then
to the A/D converter LabPro, and finally to a computer via USB.
We use software which comes with LabPro
to control sampling frequency (integration time) and duration of recording.

Telescope pointing adjustment is the next step before the experiment.
We prepare a table of the Sun's azimuthal and
elevation angles as a function of time (e.g., at 10min interval) using an on-line tool
provided by the U.S. Naval Observatory (http://aa.usno.navy.mil/data/docs/AltAz.php).
The antenna is set to the single-dish mode (i.e., dish facing toward the Sun).
We align the planes of the mount's top plate and ladder parallel to sunlight
using their shadows.
The azimuth is set to that of the Sun, and we adjust the elevation angle of the dish to maximize
the signal from the Sun on the voltage meter.
[Our dish is an off-axis paraboloid antenna, and the direction
of the dish looks very offset from the direction of the Sun.
We therefore need to use the voltage meter.
We later installed a foot-long rod on the dish and marked a point (on the dish)
at which the shadow of the rod tip falls when pointed toward the Sun.]
We then flip the dish by $180\deg$ around the metal pole for interferometer measurements.

The signal amplitudes from the two side mirrors need to be balanced.
We check the voltage readout from each side mirror separately by blocking
the optical path of the other (or by removing the other mirror). We move the central mirror toward the side of
stronger signal to decrease its effective surface area.


\subsection{Measurements}

Once the mirrors are set and the telescope is pointed toward the Sun, we start interferometer
measurements. We should see fringes from the Sun (e.g., Figure \ref{fig:sun})
as we slew the telescope and sweep across the Sun in the azimuthal direction.
We typically spend 10-30 second on each "sweep" observation,
and then correct the telescope pointing before the next sweep.
The pattern may be seen as variations of the voltage readout,
or as a fringe pattern in a plot (Figure \ref{fig:sun}), if the LabPro and computer are already started.
The LabPro and the computer do not know about telescope pointing
and record only the readout voltage as a function of time.
We therefore need to convert the time to azimuthal angle after the measurements.
We record the start and end azimuthal angles in sweeping the Sun -- we start from
a far-off position, say 10-20$\deg$ away in azimuth, and sweep the Sun in azimuth.
We assume that the telescope slew speed is constant (approximately correct when
we record for a long time, e.g. 20-30 seconds).
The projection effect, i.e., the cos(elevation) term, must be accounted for in calculation
of arc length in the sky.

We change baseline length by sliding the side mirrors on the ladder and repeat fringe measurements.
The baseline length should be determined from the fringe pattern, but for reference, we record the side
mirror separation using a tape measure fixed to the ladder.

\subsection{Miscellaneous}

Radio interference was initially a problem.
We conducted a site search across the campus.
We brought the dish and a commercial receiver (called a satellite finder $\sim$ \$10-20, 
which is used to find commercial television satellites when a dish is installed) and
compared the strengths of the Sun and ambient radio signals.
We conveniently found that one spot in front of our building was radio quiet.

Geosynchronous satellites are located along a thin belt in the sky.
The Sun's sidereal path gets aligned along this belt in some seasons,
which hinders the experiment. This should be checked at the planning stage
of the experiment.

The current mount structure is slightly wider than a standard doorway.
It does not fit on most of our elevators and cannot
pass through exit doors of our building.
We have to carry it out via a loading deck.
This could have been taken into account when the telescope was designed.

The telescope can be used as a single-dish radio telescope by pointing the dish directly toward
the sky. The beam size of our dish is roughly $\sim 1\deg$ in X band, with which we can
barely resolve the Sun ($\sim 1/2\deg$ diameter).
We can compare the profiles of the Sun and a commercial satellite (a point source)
to find this experimentally.
The Sun's diameter can be resolved and determined with the interferometer.
The comparison of the single-dish and interferometer measurements permits students
to appreciate the superiority of interferometry in terms of spatial resolution.


\section{Results from A Lab Report}\label{sec:results}

\begin{figure}
\epsscale{1.1}
\plotone{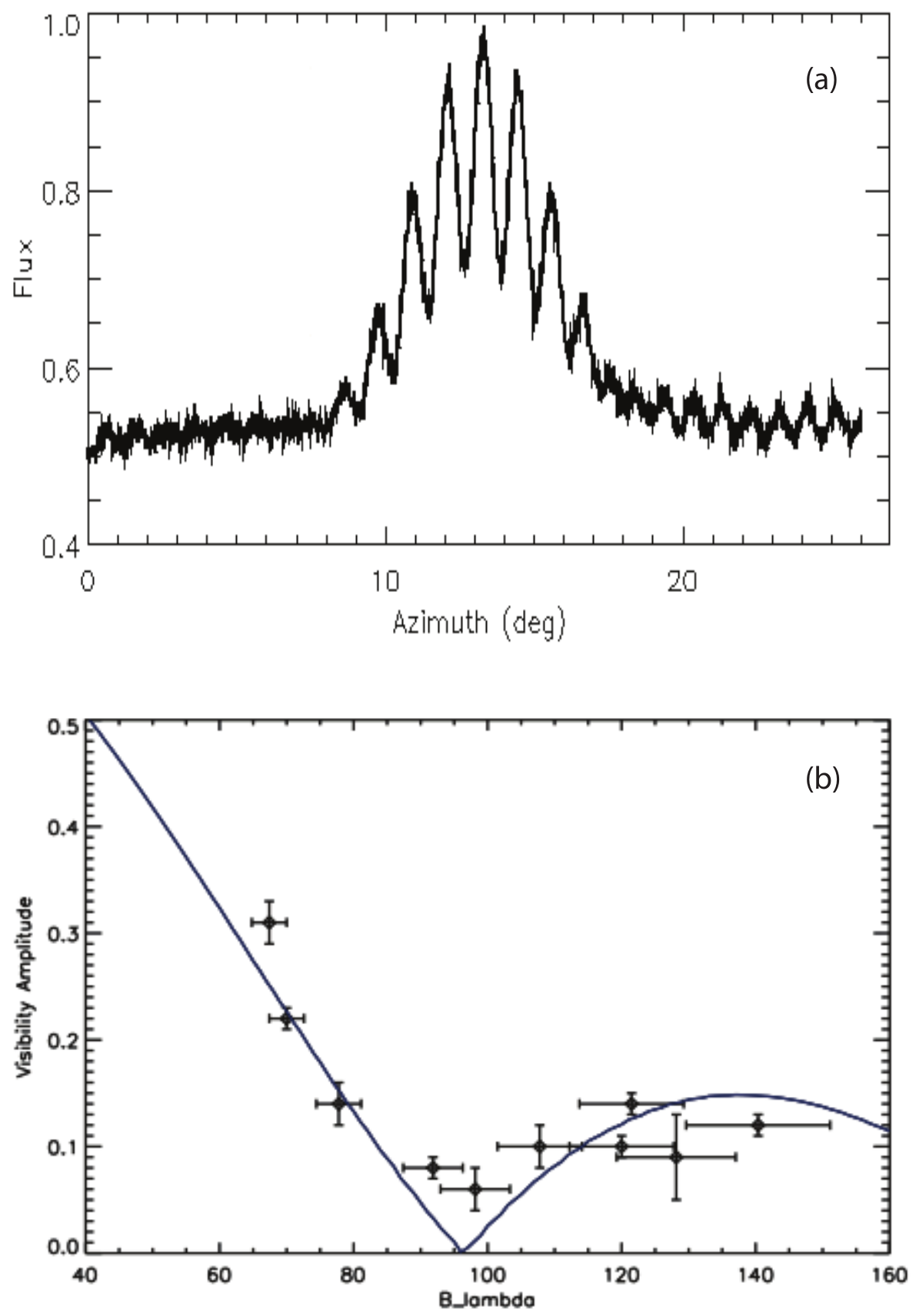}
\caption{Results from a student report.
(a) Measured fringes from the Sun.
(b) Visibility amplitude vs baseline length [in $\lambda$]. }
\label{fig:results}
\end{figure}

Figure \ref{fig:results} shows results from a student group's lab report,
Panel (a) is an example of a fringe
pattern of the Sun. They determined the baseline length by measuring the interval between
peaks and troughs (and from their readings of the side mirror separation). 
This group repeated fringe measurements three times at each of 10 different baseline lengths.
Panel (b) shows a fit of the {\it sinc} function, i.e., the Fourier transform of the Sun.
The null point at $B_{\lambda}=96$ in the fit suggests that their measurement of the sun's
diameter is $\sim 36^\prime$ at $\sim 11$ GHz.
Note that its reported diameter at $\sim 10$ GHz is about $34^\prime$ with little dependence
on solar activity (i.e., sunspot number); this diameter is calculated from
the observed radio-to-optical diameter ratios \citep{Das2000} and  the optical diameter of $\sim 30^\prime$.
These results demonstrate a proof of concept demonstrated by our students,
and a variety of exercises can be developed for a student lab beyond what is described here.


\acknowledgments

We thank Peter Koch, the previous Chair of the Department of Physics and Astronomy at Stony Brook University,
for providing funds to develop this experiment.
We also thank Munetake Momose for useful discussions.
We also thank students in the lab course, Kendra Kellogg, Melissa Louie, and Stephanie Zajac, for letting us
use plots from their lab report.
This work is supported by the NSF through grant AST-1211680. JK also acknowledges the supports from NASA
through grants NNX09AF40G, NNX14AF74G, a Herschel Space Observatory grant, and a Hubble Space Telescope grant.




\end{document}